\documentclass[twoside,11pt]{article}

\usepackage{amsmath,amssymb,amsfonts,amsthm,amsbsy}
\usepackage[utf8]{inputenc}
\usepackage{upgreek}
\usepackage{hyperref}
\usepackage{enumerate}
\usepackage{graphicx}
\usepackage{color}
\usepackage{sectsty}
\usepackage{tocloft}
\usepackage[margin=1.25in]{geometry}
\usepackage[usenames,dvipsnames,svgnames,table]{xcolor}

\definecolor{labelcolor}{RGB}{100,0,0}

\partfont{\Large\bfseries\centering}

\begin{document}
\title{Chaotic dynamics with Maxima}
\author{A. Morante and J. A. Vallejo\\
{\normalsize Facultad de Ciencias}\\
{\normalsize Universidad Aut\'onoma de San Luis Potos\'i (M\'exico)}\\
{\normalsize Av. Salvador Nava s/n Col. Lomas CP 78290 San Luis Potos\'i (SLP)}\\
{\footnotesize Email: \texttt{amorante,jvallejo@fciencias.uaslp.mx}}
}
\date{\today}
\maketitle

\begin{abstract}
We present an introduction to the study of chaos in discrete and continuous
dynamical systems using the CAS Maxima. These notes are intended to cover the
standard topics and techniques: discrete and continuous 
logistic equation to model growth population, staircase plots, bifurcation
diagrams and chaos transition, nonlinear continuous dynamics (Lorentz system
and Duffing oscillator), Lyapunov exponents, Poincar\'e sections, fractal 
dimension and strange attractors. The distinctive feature here is the use of 
free software with just one ingredient: the CAS Maxima. It is cross-platform
and have extensive on-line documentation.
\end{abstract}
\tableofcontents

\clearpage

\section{Introduction}
The use of a computer is a natural way to explore those notions related to dynamical systems; here, we
present the basics of chaotic dynamics using Maxima. This is a free, cross-platform, general
purpose computer algebra system \cite{Max 11}, capable also of doing numerical computations,
that we have used in short introductory summer courses (about 10 hours).
The following notes are the outcomes of such courses, and given the target audience, 
we have made every effort to keep them at an elementary level, relegating the most technical points to
the footnotes.

Of course, it is impossible to include all topics one would wish, so we have excluded those 
theoretical results which are hard to implement using the computer, because they involve irrational numbers.
Explicitly, we only make some brief remarks about results such as the density of periodic orbits or sensitivity
to initial conditions and leave aside topological transitivity.

In writing this paper, we have used Maxima version 5.24.0 and its graphical interface wxMaxima version
11.04.0. The version of Gnuplot was 4.4 patchlevel 3, everything running on a generic desktop PC with Slackware Linux 13.1.

\section{The logistic equation: continuous case}

The simplest model of population growth is given by:
\begin{equation}\label{eq1}
                      \frac{\mathrm{d}P(t)}{\mathrm{d}t} =k\cdot P(t)
\end{equation}
where $P(t)$ is the population at time $t$ and $k$ is a constant, positive for an increasing
population and negative for a decreasing one. However, in the long run this is not a good
model: it disregards limiting factors such as propagations of diseases or
lacks of food supply. A simple modification in \eqref{eq1}
which takes into account these factors can be obtained replacing $k$ with a parameter $K=K(P)$, such
that it decreases when $P$ increases. A possible model for this, is $K=a-b\cdot P$, $a\gg b>0$, so
\eqref{eq1} takes the form
\begin{equation}\label{eq3}
                 \frac{\mathrm{d}P(t)}{\mathrm{d}t} = (a-b\cdot P(t)) \cdot P(t).
\end{equation}
If $P(t)\simeq 0$, equation \eqref{eq3} becomes equation \eqref{eq1}, with $k\simeq a$,
but for increasing values of $P(t)$, $b\cdot P(t)$ approach $a$, and therefore
$\frac{\mathrm{d}P(t)}{\mathrm{d}t}\simeq 0$, slowing down the population's growth.
Making the change of variables
\begin{equation}\label{eq4}
P=\frac{ax}{b},\quad  \frac{\mathrm{d}P(t)}{\mathrm{d}t}=\frac{a}{b}\frac{\mathrm{d}x}{\mathrm{d}t}
\end{equation}
equation \eqref{eq3} becomes (with $a=k$)
\begin{equation}\label{eq5}
                        \frac{\mathrm{d}x}{\mathrm{d}t} =k\cdot x\cdot (1-x),
\end{equation}
which is called the logistic equation (it was introduced by P. F. Verhulst in $1838$,
a discrete version was popularized by R. May in the seventies, see \cite{May 76}).

We will use Maxima to solve equation \eqref{eq5} (note the use of the apostrophe to define an equation. The command for solving first \emph{or} second-order ODEs is \texttt{ode2}):

\noindent
\begin{minipage}[t]{8ex}{\color{red}\bf
\begin{verbatim}
(%i1) 
\end{verbatim}}
\end{minipage}
\begin{minipage}[t]{\textwidth}{\color{blue}
\begin{verbatim}
'diff(x,t)=k*x*(1-x);
\end{verbatim}}
\end{minipage}
\begin{math}\displaystyle
\parbox{8ex}{\color{labelcolor}(\%o1) }
\frac{d}{d\,t}\,x=k\,\left( 1-x\right) \,x
\end{math}

\noindent
\begin{minipage}[t]{8ex}{\color{red}\bf
\begin{verbatim}
(%i2) 
\end{verbatim}}
\end{minipage}
\begin{minipage}[t]{\textwidth}{\color{blue}
\begin{verbatim}
ode2(%,x,t);
\end{verbatim}}
\end{minipage}
\begin{math}\displaystyle
\parbox{8ex}{\color{labelcolor}(\%o2) }
\frac{\mathrm{log}\left( x\right) -\mathrm{log}\left( x-1\right) }{k}=t+\%c
\end{math}\\

\noindent Now, we set an initial condition for this \emph{first}-order (hence the 1 in \texttt{ic1}) equation $x(t=0)=x_0$:

\noindent
\begin{minipage}[t]{8ex}{\color{red}\bf
\begin{verbatim}
(%i3) 
\end{verbatim}}
\end{minipage}
\begin{minipage}[t]{\textwidth}{\color{blue}
\begin{verbatim}
ic1(%,t=0,x=x[0]),logcontract;
\end{verbatim}}
\end{minipage}
\begin{math}\displaystyle
\parbox{8ex}{\color{labelcolor}(\%o3) }
\frac{\mathrm{log}\left( \frac{x}{x-1}\right) }{k}=\frac{k\,t+\mathrm{log}\left( \frac{{x}_{0}}{{x}_{0}-1}\right) }{k}
\end{math}\\

\noindent and solving the last equation we get

\noindent
\begin{minipage}[t]{8ex}{\color{red}\bf
\begin{verbatim}
(%i4) 
\end{verbatim}}
\end{minipage}
\begin{minipage}[t]{\textwidth}{\color{blue}
\begin{verbatim}
solve(%,x);
\end{verbatim}}
\end{minipage}
\begin{math}\displaystyle
\parbox{8ex}{\color{labelcolor}(\%o4) }
[x=\frac{{x}_{0}\,{e}^{k\,t}}{{x}_{0}\,{e}^{k\,t}-{x}_{0}+1}]
\end{math}\\

\noindent Thus, we have the explicit solution (note that the outcome (\%o4) is
a \emph{list} --a collection of elements enclosed between brackets-- and we want to define our solution as
the right-hand side of its first element):

\noindent
\begin{minipage}[t]{8ex}{\color{red}\bf
\begin{verbatim}
(%i5) 
\end{verbatim}}
\end{minipage}
\begin{minipage}[t]{\textwidth}{\color{blue}
\begin{verbatim}
define(x(t),rhs(first(%o4)))$
\end{verbatim}}
\end{minipage}

Let us note that normalization \eqref{eq4} implies that for all initial conditions
$0<x_0$, the solutions $x(t)\to 1$ when $t \to \infty$. In fact, even if we had not known
the solutions explicitly, a simple continuity argument would show that all
solutions bounded by the steady states $x=0$ and $x=1$, tend monotonically to $x=1$
\footnote{For a system $\dot{x}=f(x)$, in the region between two consecutive
steady states $x(t)=x^*$ with $f(x^* )=0$, $x^*_1 < x^*_2$, the mapping $f$ (and therefore
the derivative $\dot{x}$) has constant sign, so a solution $x(t)$ with initial condition
$x_0$ such that $x^*_1< x_0 <x^*_2$, is strictly monotonic. This solution is bounded, so
$\lim_{t\to \infty}x(t) = x^*_2$ if $x(t)$ is increasing, and 
$\lim_{t\to \infty}x(t) = x^*_1$ if $x(t)$ is decreasing. Finally, because of the uniqueness
theorem for the Cauchy problem,  $x(t)$ can not reach the values $x^*_1 ,x^*_2$. A similar reasoning
can be used for unbounded regions.}.
Also, in the unbounded region $x>1$ all solutions tend monotonically to the steady state $x=1$.
For $x<0$ there is no solution (populations are non negative).
In general, in a one dimensional continuous system, solutions either converge monotonically  to a steady state or diverge.

The parameter $k$ affects the slope of the solutions, as we can see by graphing them together
for different values of $k$ (in this example, varying from $k=0{.}25$ to $k=1{.}75$ with step $0{.}25$).

\noindent
\begin{minipage}[t]{8ex}{\color{red}\bf
\begin{verbatim}
(%i6) 
\end{verbatim}}
\end{minipage}
\begin{minipage}[t]{\textwidth}{\color{blue}
\begin{verbatim}
wxplot2d(makelist(subst([k=d*0.25,x[0]=0.1],x(t)),d,1,7),
[t,0,15],cons(legend,makelist(k=d*0.25,d,1,7)),
[gnuplot_preamble,"set key right bottom"]);
\end{verbatim}}
\end{minipage}
\begin{math}\displaystyle
\parbox{8ex}{\color{labelcolor}(\%t6) }
\includegraphics[width=8cm]{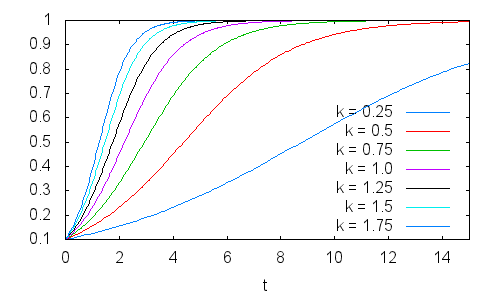}
\end{math}
\begin{math}\displaystyle
\parbox{8ex}{\color{labelcolor}(\%o6) }
\end{math}

\section{The logistic equation: discrete case}

The discrete version of equation \eqref{eq5} is the difference equation (with $r>0$)
\begin{equation}\label{eq6}
                          x_{n+1}=r\cdot x_n\cdot (1-x_n).
\end{equation}
In general, equation \eqref{eq6} can't be solved exactly\footnote{This sentence deserves a clarification.
Actually, ``closed'' solutions do exist, although defined throuh implicit functional formulas which are
`hardly useful for computational purposes', see \cite{Bru 98} and references therein. However, there are
well-known explicit solutions \emph{for particular values} of $k$, see  \url{http://en.wikipedia.org/wiki/Logistic\_map\#Solution\_in\_some_cases}.}.
Let us define the so called \emph{logistic map} (in fact, a family of \emph{quadratic maps}),
$F(r,x)=r\cdot x\cdot (1-x)$.  From \eqref{eq6} we see that, for a given $r$,
the fixed points of  $F(r,x)$ are the only constant solutions $x_n = c$ of \eqref{eq6}. 
These \emph{steady states} can be easily calculated:

\noindent
\begin{minipage}[t]{8ex}{\color{red}\bf
\begin{verbatim}
(%i7) 
\end{verbatim}}
\end{minipage}
\begin{minipage}[t]{\textwidth}{\color{blue}
\begin{verbatim}
solve(r*c*(1-c)=c,c);
\end{verbatim}}
\end{minipage}
\begin{math}\displaystyle
\parbox{8ex}{\color{labelcolor}(\%o7) }
[c=\frac{r-1}{r},c=0]
\end{math}\\

From our experience with the continuous case, we are tempted to say that
any solution of the logistic map is either a sequence bounded by the
constant solutions $x_n =0$ and $x_n =(r-1)/r$ (asymptotically approaching one of these),
or a divergent one. But this is not true: The behavior of discrete systems is quite different from that
of their continuous counterparts. To be sure, we make a graphical analysis of the solutions.
First, we define the \emph{evolution operator} of the system: 

\noindent
\begin{minipage}[t]{8ex}{\color{red}\bf
\begin{verbatim}
(%i8) 
\end{verbatim}}
\end{minipage}
\begin{minipage}[t]{\textwidth}{\color{blue}
\begin{verbatim}
F(r,x):=r*x*(1-x)$
\end{verbatim}}
\end{minipage}\\

Next, we use the \texttt{evolution} command (included in the \texttt{dynamics} package) to
calculate, for a fixed $r$, the $n+1$ points $x_{i+1}=F(r,x_i)$ from $i=0$ to $i=n$, where the
initial value $x_0$ is given (here we use a pseudo-random initial condition between $0$ and $1$).

This set of points is a segment of the \emph{orbit of $x_0$}. We start with $r=0{.}25$ making
$15$ iterations:

\noindent
\begin{minipage}[t]{8ex}{\color{red}\bf
\begin{verbatim}
(%i9) 
\end{verbatim}}
\end{minipage}
\begin{minipage}[t]{\textwidth}{\color{blue}
\begin{verbatim}
load(dynamics)$
\end{verbatim}}
\end{minipage}

\noindent
\begin{minipage}[t]{8ex}{\color{red}\bf
\begin{verbatim}
(%i10) 
\end{verbatim}}
\end{minipage}
\begin{minipage}[t]{\textwidth}{\color{blue}
\begin{verbatim}
set_random_state(make_random_state (true))$
\end{verbatim}}
\end{minipage}

\noindent
\begin{minipage}[t]{8ex}{\color{red}\bf
\begin{verbatim}
(%i11) 
\end{verbatim}}
\end{minipage}
\begin{minipage}[t]{\textwidth}{\color{blue}
\begin{verbatim}
x[0]:random(1.0);
\end{verbatim}}
\end{minipage}
\definecolor{labelcolor}{RGB}{100,0,0}
\begin{math}\displaystyle
\parbox{8ex}{\color{labelcolor}(\%o11) }
{.}8886589561406515
\end{math}

\noindent
\begin{minipage}[t]{8ex}{\color{red}\bf
\begin{verbatim}
(%i12) 
\end{verbatim}}
\end{minipage}
\begin{minipage}[t]{\textwidth}{\color{blue}
\begin{verbatim}
evolution(F(0.25,x),x[0],15,[y,0,1]);
\end{verbatim}}
\end{minipage}
\begin{math}\displaystyle
\parbox{8ex}{\color{labelcolor}(\%t12) }
\includegraphics[width=8cm]{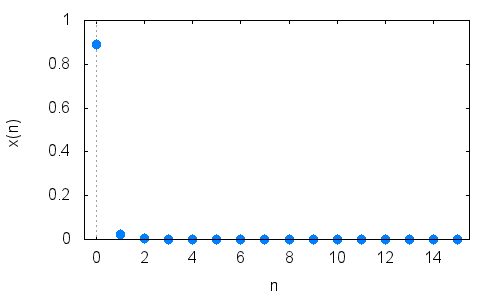}
\end{math}
\begin{math}\displaystyle
\parbox{8ex}{\color{labelcolor}(\%o12) }
\end{math}\\

In this case, the population dies. Indeed, this unlucky fate is independent
of the initial condition $x_0$, whenever $0<r\leq 1$ and $x_0 \in \left] \frac{r-1}{r},\frac{1}{r}\right[$
(let us note that this last condition is always fulfilled if $x_0 \in ] 0,1 [$).
For $0<r\leq 1$, this parameter controls the rate at which population dies, as the reader can
check by using other values for $r$, such as $r=0{.}123456789$.

A few experiments will lead us to the following rules for the behavior of the solutions:
\begin{itemize}
\item If $0<r\leq 1$ and $x_0 \in \left] \frac{r-1}{r},\frac{1}{r}\right[$, the population
    eventually dies, independently of the initial condition.
\item If $1<r\leq 2$ and $x_0 \in ]0,1[$, the population quickly approaches the value $(r-1)/r$, 
      independently of the initial condition.
\item If $2<r\leq 3$ and $x_0 \in ]0,1[$, the population tends again to the value
      $(r-1)/r$, but in an oscillating way (maybe after a short transient). The rate of
      convergence is linear, except for $r=3$, where the rate of convergence is quite slow,
      in fact it is sub-linear.
\item If $3<r<1+\sqrt{6}$ (with $1+\sqrt{6}\simeq 3{.}45$), the population oscillates between
      two values, almost independently of the initial condition. These two values, which depend on
      $r$, are said to have \emph{primary period two}.
\item If $3{.}45<r<3{.}54$ (approximately), for almost all initial condition the population oscillates
      between four periodic points.
\item If $r$ increases over $3{.}54$, for almost all initial condition the population oscillates
      between $8$ periodic points, then $16$, $32$, etc. The size of the intervals formed by the values
      of the parameter producing oscillations of a given length, becomes small, and the quotient
      of the size for two consecutive period-doubling intervals approaches the so called
      \emph{Feigenbaum constant} $F=4{.}669$... This behavior is called a \emph{period-doubling cascade}.
\end{itemize}

Let us note that when the parameter $r$ increases, the dynamics differs from that observed
in the continuous case.

Let us show some examples. For our initial pseudo-random $x_0$ and $r=1{.}3$:

\noindent
\begin{minipage}[t]{8ex}{\color{red}\bf
\begin{verbatim}
(%i13) 
\end{verbatim}}
\end{minipage}
\begin{minipage}[t]{\textwidth}{\color{blue}
\begin{verbatim}
evolution(F(1.3,x),x[0],15,[y,0,1]);
\end{verbatim}}
\end{minipage}
\begin{math}\displaystyle
\parbox{8ex}{\color{labelcolor}(\%t13) }
\includegraphics[width=8cm]{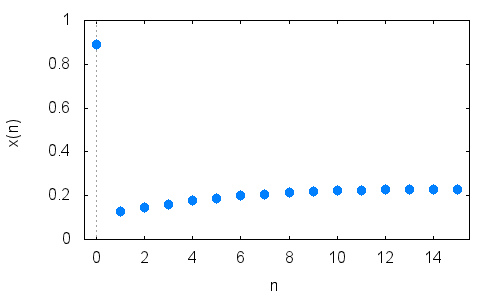}
\end{math}
\begin{math}\displaystyle
\parbox{8ex}{\color{labelcolor}(\%o13) }
\end{math}\\

We note that $(1{.}3-1)/1{.}3 \simeq 0{.}23$.
For $r=3{.}5$ and the initial condition $x_0 =0.3$, we observe the
transient and four periodic points:

\noindent
\begin{minipage}[t]{8ex}{\color{red}\bf
\begin{verbatim}
(%i14) 
\end{verbatim}}
\end{minipage}
\begin{minipage}[t]{\textwidth}{\color{blue}
\begin{verbatim}
evolution(F(3.5,x),0.3,25,[y,0,1]);
\end{verbatim}}
\end{minipage}
\begin{math}\displaystyle
\parbox{8ex}{\color{labelcolor}(\%t14) }
\includegraphics[width=8cm]{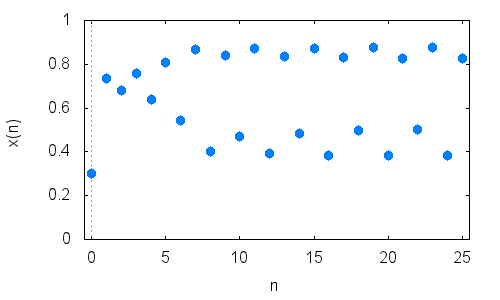}
\end{math}
\begin{math}\displaystyle
\parbox{8ex}{\color{labelcolor}(\%o14) }
\end{math}

\section{Staircase diagrams}

There is another graphical method to study the phenomenon of the emergence of ``attracting'' periodic points that we have seen in the last example, based on the particular form of the evolution equation:
$
                                x_{n+1}=f(x_n)
$.
For the logistic map ($f(x)=r\cdot x\cdot (1-x)$), if the coordinates of the
points in the plane represent two consecutive values in the orbit of $x_0$, the point
$(x,y) = (x_n,x_{n+1})$ can be obtained graphically as the intersection of the vertical line
$x=x_n$ and the graph of the function $y=f(x)$. Once $x_{n+1}$ is known, in order to get
$x_{n+2}$ we just have to let $x_{n+1}$ play the r\^ole of $x_n$ in the previous step. So, we take
the intersection of the horizontal line $y=x_{n+1}$ with the diagonal $y=x$. Now
$x_{n+2}=f(x_{n+1})$, and by iterating the process a predetermined number of times, we 
get a segment of the orbit of $x_0$. Maxima implements this construction with the \texttt{staircase}
command:

\noindent
\begin{minipage}[t]{8ex}{\color{red}\bf
\begin{verbatim}
(%i15) 
\end{verbatim}}
\end{minipage}
\begin{minipage}[t]{\textwidth}{\color{blue}
\begin{verbatim}
staircase(F(3.1,x),x[0],250,[x,0,1],[y,0,1]);
\end{verbatim}}
\end{minipage}
\begin{math}\displaystyle
\parbox{8ex}{\color{labelcolor}(\%t15) }
\includegraphics[width=8cm]{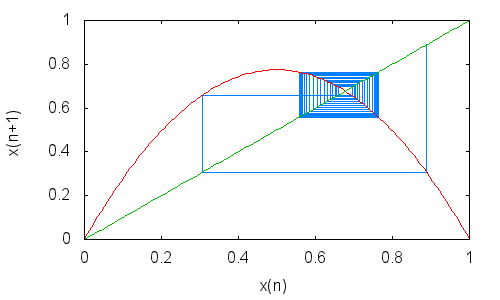}
\end{math}
\begin{math}\displaystyle
\parbox{8ex}{\color{labelcolor}(\%o15) }
\end{math}\\

Here we see the orbit oscillating between the periodic points ($x\simeq 0{.}56$) and
($x\simeq 0{.}76$). We can also reproduce the behavior of (\%o14),
where there is a period-four orbit (we make $500$ iterations to get a distinguishable path). Let us note
that each ``rectangle'' intersecting the diagonal defines two periodic points:

\noindent
\begin{minipage}[t]{8ex}{\color{red}\bf
\begin{verbatim}
(%i16) 
\end{verbatim}}
\end{minipage}
\begin{minipage}[t]{\textwidth}{\color{blue}
\begin{verbatim}
staircase(F(3.5,x),x[0],500,[x,0,1],[y,0,1]);
\end{verbatim}}
\end{minipage}
\begin{math}\displaystyle
\parbox{8ex}{\color{labelcolor}(\%t16) }
\includegraphics[width=8cm]{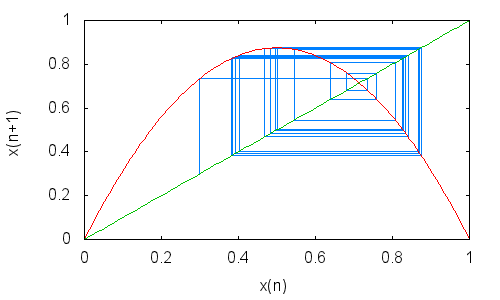}
\end{math}
\begin{math}\displaystyle
\parbox{8ex}{\color{labelcolor}(\%o16) }
\end{math}

\section{Bifurcations and chaos}

We have seen that the behavior of the orbits for the logistic map depends on the parameter $r$.
In fact, from $r=3$ onwards, the number of periodic points increases in a kind of
``period-doubling cascade'' (see (\%o17) below). It is possible to study the change in the structure
of the periodic orbits using the so called \emph{bifurcation diagrams}. In these, we
represent the values of the parameter $r$ on the horizontal axis. For each of these values we
mark the values of the corresponding ``attracting'' periodic points
(the use of attracting points is a technical issue, that we will not address here). Maxima implements
these plots in its built-in \texttt{orbits} command.
In the example below, we restrict the plot to the iterations between $n=150$ and $n=200$,
for a given initial condition $x_0$. The range of the values of $r$ is $[2{.}5, 4]$:

\noindent
\begin{minipage}[t]{8ex}{\color{red}\bf
\begin{verbatim}
(%i17) 
\end{verbatim}}
\end{minipage}
\begin{minipage}[t]{\textwidth}{\color{blue}
\begin{verbatim}
orbits(F(r,x), x[0], 150, 200, [r, 2.5, 4], [style, dots]);
\end{verbatim}}
\end{minipage}
\begin{math}\displaystyle
\parbox{8ex}{\color{labelcolor}(\%t17) }
\includegraphics[width=8cm]{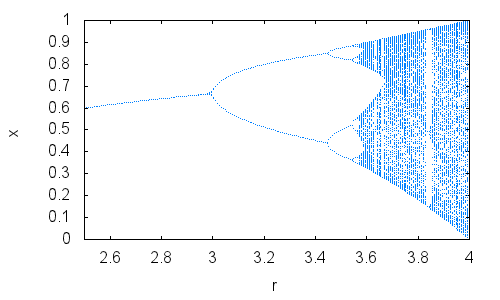}
\end{math}
\begin{math}\displaystyle
\parbox{8ex}{\color{labelcolor}(\%o17) }
\end{math}\\

The diagram so obtained displays a special property: it is \emph{self-similar}. This means that after a
change of scale, the resulting picture has the same structure that the original one. For example,
if we magnify (\%o17) centering the zoom on the second bifurcation of the lower branch, we get:

\noindent
\begin{minipage}[t]{8ex}{\color{red}\bf
\begin{verbatim}
(%i18) 
\end{verbatim}}
\end{minipage}
\begin{minipage}[t]{\textwidth}{\color{blue}
\begin{verbatim}
orbits(F(r,x),x[0],150,200,[r,3.5,3.6],[x,0.3,0.4],
[style,dots]);
\end{verbatim}}
\end{minipage}
\begin{math}\displaystyle
\parbox{8ex}{\color{labelcolor}(\%t18) }
\includegraphics[width=8cm]{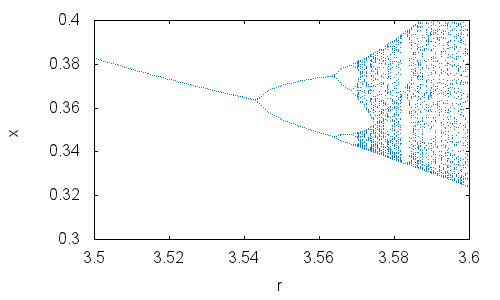}
\end{math}
\begin{math}\displaystyle
\parbox{8ex}{\color{labelcolor}(\%o18) }
\end{math}\\

A measure of the self-similarity of a set is its \emph{fractal dimension} which, unlike the
dimension of a vector space, can be a non integer number. In Section \ref{fractals} we show an
example of a set having non integer fractal dimension.

The period-doubling cascade and the appearance of sets with fractal dimension, are usually signs
of chaotic behavior. A dynamical system with whose evolution is described by a continuous function
$f:I\subset\mathbb{R}\to I$ is said to be \emph{chaotic}\footnote{For simplicity we consider only 
one-dimensional systems, but the case of multidimensional systems is analogous. For further details
on the chaos definition see \cite{Ban 92}, \cite{Dev 89}, \cite{VB 94}} if:
\begin{enumerate}[(a)]
\item It has \emph{sensitivity to initial conditions}: a small variation in the initial
  condition $x_0$ of an orbit can produce, in the long run, another orbit far away from the
  original one.
\item \emph{Periodic points are ``dense''} in the phase space $I$ of the system: in general, a set
 $S$ is dense in the set $I\subset \mathbb{R}$ if for any point $p \in I$
 and any $\varepsilon >0$, the interval $]p-\varepsilon,p+\varepsilon[$ intersects $S$.
\item It has the \emph{mixing property} (we also say that the system is
  \emph{topologically transitive}): for any two intervals $J,K \subset I$ there exist points
  of $J$ whose orbits eventually enter $K$, that is, there exists an $n>1$ such that 
  \[
    f_n (J)\cap K \neq \emptyset ,
  \]
with $f_n =f\circ \overset{n)}{\cdots} \circ f$. This property is satisfied if and only if the system has an
orbit $\{ f_n (x_0):n\in \mathbb{N}\cup \{ 0\}\}$ that is dense in $I$. \footnote{The equivalence
is as follows: if the orbit of $x_0$ is dense in $I$, there exists an infinity of points
$x_n=f_n (x_0)$ in \emph{any} pair of intervals $J,K$, so $f$ is transitive. Conversely, if
$f$ is transitive, for a given interval $J$ let $\{ A_j \}_{j\in\mathbb{N}}$ be the basis of
open intervals with rational end points that are contained in $J$; from the transitivity of $f$
the family of open sets $B_j =\cup_{p\in\mathbb{N}}f_{-p}(A_n)$ is dense in $J$ (since, if
$U\subset J$, there exists a number $p$ such that $f_p (U)\cap A_n \neq \emptyset$, that is,
$U\cap f_p (A_n)=U\cap B_n \neq \emptyset$). From the Baire theorem, $\cap_{j\in\mathbb{N}}B_j$
is dense in $J$ too. But this intersection consists of points with dense
orbits, since if  $x\in \cap_{j\in\mathbb{N}}B_j$, for each $n\in\mathbb{N}$ there is a
$p_n \in\mathbb{N}$ such that $f_{p_n}\in A_n$. The sequence $\{f_{p_n}\}_{n\in\mathbb{N}}$
has points in each $A_n$ and therefore is dense.}
\end{enumerate}

We illustrate each one of these conditions using the logistic map in the examples below.

\begin{enumerate}[(a)]
\item Sensitivity to initial conditions.
For $r=4$, we consider the initial conditions $x_0 =\frac{r-1}{r}=\frac{3}{4}$ and
$x_0 +\varepsilon$, comparing  the respective orbits after $50$ iterations.

\begin{minipage}[t]{8ex}{\color{red}\bf
\begin{verbatim}
(%i19) 
\end{verbatim}}
\end{minipage}
\begin{minipage}[t]{\textwidth}{\color{blue}
\begin{verbatim}
evolution(F(4,x),3/4,50,[style,[lines,2]]);
\end{verbatim}}
\end{minipage}

\begin{math}\displaystyle
\parbox{8ex}{\color{labelcolor}(\%t19) }
\includegraphics[width=7cm]{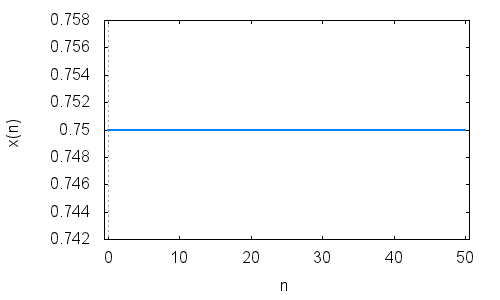}
\end{math}
\begin{math}\displaystyle
\parbox{8ex}{\color{labelcolor}(\%o19) }
\end{math}\\
\noindent Choosing $\varepsilon$ small enough (pseudo-random) we see how drastically these orbits diverge 
(we will see how to measure this divergence in section \ref{explia}):

\noindent
\begin{minipage}[t]{8ex}{\color{red}\bf
\begin{verbatim}
(%i20) 
\end{verbatim}}
\end{minipage}
\begin{minipage}[t]{\textwidth}{\color{blue}
\begin{verbatim}
eps:random(0.000000000001);
\end{verbatim}}
\end{minipage}
\definecolor{labelcolor}{RGB}{100,0,0}
\begin{math}\displaystyle
\parbox{8ex}{\color{labelcolor}(\%o20) }
9{.}87415729237455\,{10}^{-13}
\end{math}\\
\begin{minipage}[t]{8ex}{\color{red}\bf
\begin{verbatim}
(%i21) 
\end{verbatim}}
\end{minipage}
\begin{minipage}[t]{\textwidth}{\color{blue}
\begin{verbatim}
evolution(F(4,x),3/4+eps,50,[style,[lines,2]]);
\end{verbatim}}
\end{minipage}
\begin{math}\displaystyle
\parbox{8ex}{\color{labelcolor}(\%t21) }
\includegraphics[width=7cm]{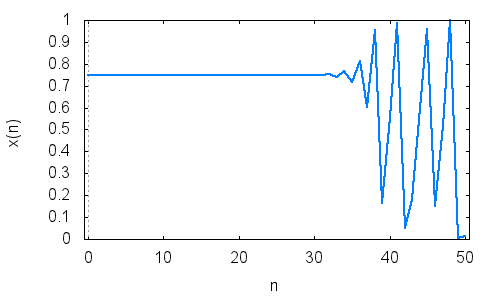}
\end{math}
\begin{math}\displaystyle
\parbox{8ex}{\color{labelcolor}(\%o21) }
\end{math}\\

\item The set of periodic points is dense in the interval $I = [0,1]$. This is clearly seen
  in (\%o17), where the right side of the unit square is ``filled'' with periodic points.

\item There exists a dense orbit. Again, we can exemplify this behavior with $k=4$ and the initial
  condition $x_0 =0{.}2$ after $50,000$ iterations. Its orbit fills the phase space:

\noindent

\begin{minipage}[t]{8ex}{\color{red}\bf
\begin{verbatim}
(%i22) 
\end{verbatim}}
\end{minipage}
\begin{minipage}[t]{\textwidth}{\color{blue}
\begin{verbatim}
evolution(F(4,x),0.2,50000,[style,dots]);
\end{verbatim}}
\end{minipage}
\definecolor{labelcolor}{RGB}{100,0,0}
\begin{math}\displaystyle
\parbox{8ex}{\color{labelcolor}(\%t22) }
\includegraphics[width=7cm]{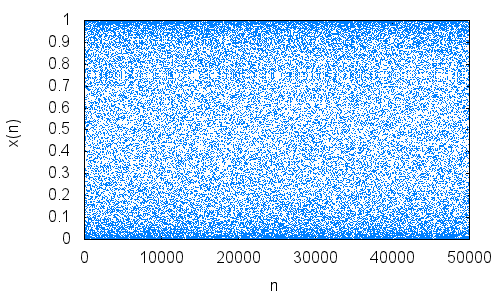}
\end{math}
\begin{math}\displaystyle
\parbox{8ex}{\color{labelcolor}(\%o22) }
\end{math}\\
\end{enumerate}

\section{The Lorenz attractor}
As mentioned, the monotonicity of the solutions restricts the presence of chaos in 
one-dimensional continuous systems. However, for higher order equations (at least of second order with
a time dependent non-homogeneous term, as in the Duffing equation shown later) or in systems of
first order equations (at least three equations, even in the case of an autonomous system, but
necessarily non linear\footnote{We have already seen that one dimensional continuous systems
can not display chaotic behavior (nor even periodic behavior). For autonomous systems in the plane
(or second order differential equations with time-independent coefficients), the uniqueness of
solutions together with the Poincar\'e-Bendixson theorem rule out the existence of chaos.
The problem of setting the ``minimal conditions'' to ensure the existence of chaos is still an open
question. An interesting report on this topic is \cite{SL 00}.}), it is possible to have chaotic behavior.
As an example, we consider the Lorenz system, the first system in which the properties
of chaotic dynamics were explicitly seen, in 1963 \cite{Lor 63}: 
\[
\begin{cases}
\dot{x} = 10y-10x \\
\dot{y} = -xz+28x-y \\
\dot{z} = xy-8z/3
\end{cases}
\]
To numerically study the orbits, we use the 
4th-order Runge-Kutta method. The following commands compute the orbit with initial
condition $(-8,8,27)$ after $50$ time units (with a step of $0{.}01$):

\noindent
\begin{minipage}[t]{8ex}{\color{red}\bf
\begin{verbatim}
(%i23) 
\end{verbatim}}
\end{minipage}
\begin{minipage}[t]{\textwidth}{\color{blue}
\begin{verbatim}
numer:false$
\end{verbatim}}
\end{minipage}

\noindent
\begin{minipage}[t]{8ex}{\color{red}\bf
\begin{verbatim}
(%i24) 
\end{verbatim}}
\end{minipage}
\begin{minipage}[t]{\textwidth}{\color{blue}
\begin{verbatim}
latractor: [10*y-10*x, -x*z+28*x-y, x*y-8*z/3]$
\end{verbatim}}
\end{minipage}

\noindent
\begin{minipage}[t]{8ex}{\color{red}\bf
\begin{verbatim}
(%i25) 
\end{verbatim}}
\end{minipage}
\begin{minipage}[t]{\textwidth}{\color{blue}
\begin{verbatim}
linitial: [-8, 8, 27]$
\end{verbatim}}
\end{minipage}

\noindent
\begin{minipage}[t]{8ex}{\color{red}\bf
\begin{verbatim}
(%i26) 
\end{verbatim}}
\end{minipage}
\begin{minipage}[t]{\textwidth}{\color{blue}
\begin{verbatim}
lsolution:rk(latractor,[x,y,z],linitial,[t,0,50,0.01])$
\end{verbatim}}
\end{minipage}

\noindent
\begin{minipage}[t]{8ex}{\color{red}\bf
\begin{verbatim}
(%i27) 
\end{verbatim}}
\end{minipage}
\begin{minipage}[t]{\textwidth}{\color{blue}
\begin{verbatim}
lpoints: map(lambda([x], rest(x)), lsolution)$
\end{verbatim}}
\end{minipage}

\noindent
\begin{minipage}[t]{8ex}{\color{red}\bf
\begin{verbatim}
(%i28) 
\end{verbatim}}
\end{minipage}
\begin{minipage}[t]{\textwidth}{\color{blue}
\begin{verbatim}
load(draw)$
\end{verbatim}}
\end{minipage}

\noindent The following command needs that the package \texttt{draw} be previously loaded (do \texttt{load(draw)} otherwise). 
Note that the output is actually a separate Gnuplot window, so (by pressing  the left button of the mouse and dragging around)
the reader can rotate the whole picture to better appreciate its details.

\noindent
\begin{minipage}[t]{8ex}{\color{red}\bf
\begin{verbatim}
(%i29) 
\end{verbatim}}
\end{minipage}
\begin{minipage}[t]{\textwidth}{\color{blue}
\begin{verbatim}
wxdraw3d(point_type=none,points_joined=true,color=orange,
xlabel="x(t)",ylabel="y(t)",zlabel="z(t)",
xtics=10,ytics=10,ztics=10,points(lpoints));
\end{verbatim}}
\end{minipage}
\begin{math}\displaystyle
\parbox{8ex}{\color{labelcolor}(\%t29) }
\includegraphics[width=9cm]{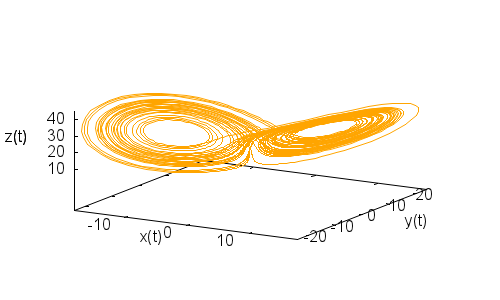}
\end{math}
\begin{math}\displaystyle
\parbox{8ex}{\color{labelcolor}(\%o29) }
\end{math}\\

The phrase ``order within chaos'' is frequently found. It means that in some systems,
even though the orbits can appear unpredictable and very complex, there is some regularity.
For example, in the Lorenz system, the sensitivity to initial conditions prevents us from making
accurate predictions, but we can assert that in the long run the orbit will be restricted to
a \emph{bounded} region of the space called an \emph{attractor}\footnote{Attractors are
classified as \emph{strange} and \emph{no strange} if they have fractal or integer dimension,
respectively. In section \ref{fractals} we will see how to compute the fractal dimension of an
attractor.}. This seems clear in figure (\%o29), but we can prove it analytically with the aid of Maxima.
Let us consider a general Lorentz system, where $\sigma, \rho$ and $\beta$ are parameters:
\begin{equation}\label{eqlorenzg}
\begin{cases}
&\dot{x} = \sigma (y-x) \\
&\dot{y} = x(\rho -z)-y \\
&\dot{z} = xy-\beta z
\end{cases},
\end{equation}

\noindent
\begin{minipage}[t]{8ex}{\color{red}\bf
\begin{verbatim}
(%i30) 
\end{verbatim}}
\end{minipage}
\begin{minipage}[t]{\textwidth}{\color{blue}
\begin{verbatim}
remfunction(x)$
\end{verbatim}}
\end{minipage}

\noindent
\begin{minipage}[t]{8ex}{\color{red}\bf
\begin{verbatim}
(%i31) 
\end{verbatim}}
\end{minipage}
\begin{minipage}[t]{\textwidth}{\color{blue}
\begin{verbatim}
gradef(x(t),%sigma*(y(t)-x(t)))$
\end{verbatim}}
\end{minipage}

\noindent
\begin{minipage}[t]{8ex}{\color{red}\bf
\begin{verbatim}
(%i32) 
\end{verbatim}}
\end{minipage}
\begin{minipage}[t]{\textwidth}{\color{blue}
\begin{verbatim}
gradef(y(t),x(t)*(%rho -z(t))-y(t))$
\end{verbatim}}
\end{minipage}

\noindent
\begin{minipage}[t]{8ex}{\color{red}\bf
\begin{verbatim}
(%i33) 
\end{verbatim}}
\end{minipage}
\begin{minipage}[t]{\textwidth}{\color{blue}
\begin{verbatim}
gradef(z(t),x(t)*y(t)-%beta *z(t))$
\end{verbatim}}
\end{minipage}\\

We consider the set of those points belonging to an orbit that, for a given instant $t$, lie on
the sphere with center $(0,0,\sigma +\rho )$ and radius $R(\sigma +\rho)$, with $R$ a constant
to be determined:

\noindent
\begin{minipage}[t]{8ex}{\color{red}\bf
\begin{verbatim}
(%i34) 
\end{verbatim}}
\end{minipage}
\begin{minipage}[t]{\textwidth}{\color{blue}
\begin{verbatim}
x(t)^2+y(t)^2+(z(t)-(%sigma +%rho))^2=R^2*(%sigma +%rho)^2;
\end{verbatim}}
\end{minipage}
\begin{math}\displaystyle
\parbox{8ex}{\color{labelcolor}(\%o34) }
{\left( \mathrm{z}\left( t\right) -\sigma-\rho\right) }^{2}+{\mathrm{y}\left( t\right) }^{2}+{\mathrm{x}\left( t\right) }^{2}={\left( \sigma+\rho\right) }^{2}\,{R}^{2}
\end{math}\\

Differentiating equation (\%o34) (using the chain rule) and substituting \eqref{eqlorenzg}, we get:

\noindent
\begin{minipage}[t]{8ex}{\color{red}\bf
\begin{verbatim}
(%i35) 
\end{verbatim}}
\end{minipage}
\begin{minipage}[t]{\textwidth}{\color{blue}
\begin{verbatim}
facsum(diff(%,t),x(t),y(t),z(t));
\end{verbatim}}
\end{minipage}
\begin{math}\displaystyle
\parbox{8ex}{\color{labelcolor}(\%o35) }
-2\,\beta\,{\mathrm{z}\left( t\right) }^{2}+2\,\beta\,\left( \sigma+\rho\right) \,\mathrm{z}\left( t\right) -2\,{\mathrm{y}\left( t\right) }^{2}-2\,\sigma\,{\mathrm{x}\left( t\right) }^{2}=0
\end{math}\\

This looks like an ellipsoid, and we can prove it is one by ``completing the squares'':

\noindent
\begin{minipage}[t]{8ex}{\color{red}\bf
\begin{verbatim}
(%i36) 
\end{verbatim}}
\end{minipage}
\begin{minipage}[t]{\textwidth}{\color{blue}
\begin{verbatim}
factor(solve((sqrt(2*%beta)*z(t)-u)^2-u^2=
2*%beta*z(t)^2-2*%beta*(%sigma+%rho)*z(t),u));
\end{verbatim}}
\end{minipage}
\begin{math}\displaystyle
\parbox{8ex}{\color{labelcolor}(\%o36) }
[u=\frac{\sqrt{\beta}\,\left( \sigma+\rho\right) }{\sqrt{2}}]
\end{math}\\

So, the points $(x(t),y(t),z(t))$ should satisfy
\[
-2\sigma x(t)^2-2y(t)^2
-2\beta \left( z(t)-\frac{\sigma +\rho}{2} \right)^2 +\beta\frac{(\sigma +\rho)^2}{2} =0,
\]
which is the equation of an ellipsoid with center $(0,0,(\sigma +\rho)/2)$. The semi-axis
along directions $\mathrm{OX}$, $\mathrm{OY}$ and $\mathrm{OZ}$ are, respectively:

\noindent
\begin{minipage}[t]{8ex}{\color{red}\bf
\begin{verbatim}
(%i37) 
\end{verbatim}}
\end{minipage}
\begin{minipage}[t]{\textwidth}{\color{blue}
\begin{verbatim}
solve(1/a = sqrt(2*%sigma/(%beta*(%sigma +%rho)^2/2)),a);
\end{verbatim}}
\end{minipage}
\begin{math}\displaystyle
\parbox{8ex}{\color{labelcolor}(\%o37) }
[a=\frac{\left| \sigma+\rho\right| }{2\,\sqrt{\frac{\sigma}{\beta}}}]
\end{math}

\noindent
\begin{minipage}[t]{8ex}{\color{red}\bf
\begin{verbatim}
(%i38) 
\end{verbatim}}
\end{minipage}
\begin{minipage}[t]{\textwidth}{\color{blue}
\begin{verbatim}
solve(1/b = sqrt(2/(%beta*(%sigma +%rho)^2/2)),b);
\end{verbatim}}
\end{minipage}
\begin{math}\displaystyle
\parbox{8ex}{\color{labelcolor}(\%o38) }
[b=\frac{\left| \sigma+\rho\right| }{2\,\sqrt{\frac{1}{\beta}}}]
\end{math}

\noindent
\begin{minipage}[t]{8ex}{\color{red}\bf
\begin{verbatim}
(%i39) 
\end{verbatim}}
\end{minipage}
\begin{minipage}[t]{\textwidth}{\color{blue}
\begin{verbatim}
solve(1/c = sqrt(2*%beta/(%beta*(%sigma +%rho)^2/2)),c);
\end{verbatim}}
\end{minipage}
\begin{math}\displaystyle
\parbox{8ex}{\color{labelcolor}(\%o39) }
[c=\frac{\left| \sigma+\rho\right| }{2}]
\end{math}\\

These lengths depend only on the parameters $\sigma$, $\rho$ and $\beta$. Thus, taking
$R$ large enough (explicitly $R>\frac{\sigma +\rho}{2}\max \{ 2,\sqrt{1+\beta},\sqrt{1+\beta/\sigma}\}$),
we obtain a sphere containing the initial conditions (here $(x(0),y(0),z(0))=(0,0,0)$ for simplicity)
and such that when they evolve in time, all orbits remain inside this sphere.\\

As already pointed, the presence of attractors having fractal structure (strange attractors)
is a signal for chaotic dynamics, but it is neither a necessary nor a sufficient condition. 
There are attractors derived from chaotic dynamics that are not strange (for example,
in the logistic equation with $k=4$ the attractor is the whole interval $[0,1]$) and there
are non-chaotic systems displaying attractors with fractal dimension (see \cite{GOPY 84}).

\section{Lyapunov exponents}\label{explia}
We have seen that a feature of chaotic dynamical systems is the sensitivity to
initial conditions. Lyapunov exponents are introduced to qualitatively measure this property. The
idea is quite simple: By choosing a fixed orbit, we compare it with other
orbits having close initial conditions, and then measure the distance between them with a factor
of the form $\exp(\lambda t)$. Here $\lambda$ is the Lyapunov exponent. If it is positive,
the orbits diverge asymptotically and there is sensitivity to initial conditions (greater when when $\lambda$ is large).
If the exponent $\lambda$ is negative, the orbits must asymptotically approximate each other and there is
no sensitivity to initial conditions.
A value of $\lambda =0$ indicates that the orbit considered is a stable fixed point.

Let us consider, for simplicity, a one-dimensional dynamical system
\[
x_{n+1}=f(x_n )
\]
with $f:\mathbb{R}\to\mathbb{R}$ derivable except for a finite number of points (this is the case,
for example, of the logistic map). We consider the orbits starting at the points $x_0$ and
$x_0 +\varepsilon$. After $N$ iterations the points on the orbits will be, respectively,
$f_N (x_0 )$ and $f_N (x_0 +\varepsilon )$, where $f_N =f\circ \cdots \circ f$. So,
after $N$ iterations the distance between the orbits is
$| f_N (x_0 +\varepsilon )-f_N (x_0 )|$, which can be written in the form 
\[
| f_N (x_0 +\varepsilon )-f_N (x_0 )|=\varepsilon e^{N\lambda (x_0 )}
\]
for a suitable $\lambda (x_0 )\in\mathbb{R}$ that is called the \emph{Lyapunov exponent} at the
point $x_0$. Is in this way that the Lyapunov exponent gives a measure of how the initial
separation $\varepsilon$ is amplified when the orbits evolve.

Solving the last equation for $\lambda (x_0)$:

\noindent
\begin{minipage}[t]{8ex}{\color{red}\bf
\begin{verbatim}
(%i40) 
\end{verbatim}}
\end{minipage}
\begin{minipage}[t]{\textwidth}{\color{blue}
\begin{verbatim}
assume(%epsilon>0,N>0)$
\end{verbatim}}
\end{minipage}

\noindent
\begin{minipage}[t]{8ex}{\color{red}\bf
\begin{verbatim}
(%i41) 
\end{verbatim}}
\end{minipage}
\begin{minipage}[t]{\textwidth}{\color{blue}
\begin{verbatim}
remvalue(x[0])$
\end{verbatim}}
\end{minipage}

\noindent
\begin{minipage}[t]{8ex}{\color{red}\bf
\begin{verbatim}
(%i42) 
\end{verbatim}}
\end{minipage}
\begin{minipage}[t]{\textwidth}{\color{blue}
\begin{verbatim}
solve(abs(f[N](x[0]+%epsilon)-f[N](x[0]))=
      %epsilon*exp(N*%lambda),%lambda);
\end{verbatim}}
\end{minipage}
\definecolor{labelcolor}{RGB}{100,0,0}
\begin{math}\displaystyle
\parbox{8ex}{\color{labelcolor}(\%o42) }
[\lambda=\frac{\mathrm{log}\left( \frac{\left| {f}_{N}\left( \epsilon+{x}_{0}\right) -{f}_{N}\left( {x}_{0}\right) \right| }{\epsilon}\right) }{N}]
\end{math}\\

\noindent and taking limits with $\varepsilon \to 0$ and $N\to \infty$ 
(taking into account the continuity of $\log$ and the definition of the derivative), we have:
\[
\lambda (x_0)=\lim_{N\to\infty}\frac{1}{N}\log \left| \frac{\mathrm{d}f_N}{\mathrm{d}x} (x_0)\right| .
\]
Just to show the symbolic capabilities of Maxima, we will use it to evaluate the derivative in the last
expression. First, we declare the derivative of the function $f$ by giving it a name (the usual $f'$):

\noindent
\begin{minipage}[t]{8ex}{\color{red}\bf
\begin{verbatim}
(%i43) 
\end{verbatim}}
\end{minipage}
\begin{minipage}[t]{\textwidth}{\color{blue}
\begin{verbatim}
gradef(f(x),f\'(x))$
\end{verbatim}}
\end{minipage}

\noindent Then, we set the values $x_1 =f(x_0)$ and $f'(x_0 )=\frac{\mathrm{d}f}{\mathrm{d}x}(x_0)$:

\noindent
\begin{minipage}[t]{8ex}{\color{red}\bf
\begin{verbatim}
(%i44) 
\end{verbatim}}
\end{minipage}
\begin{minipage}[t]{\textwidth}{\color{blue}
\begin{verbatim}
atvalue(f(y),y=x[0],x[1])$
\end{verbatim}}
\end{minipage}

\noindent
\begin{minipage}[t]{8ex}{\color{red}\bf
\begin{verbatim}
(%i45) 
\end{verbatim}}
\end{minipage}
\begin{minipage}[t]{\textwidth}{\color{blue}
\begin{verbatim}
atvalue('diff(f(y),y),y=x[0],f\'(x[0]))$
\end{verbatim}}
\end{minipage}

\noindent Now, we compute the derivative of $f_2 =f\circ f$ at $x_0$ (Maxima knows the chain rule):

\noindent
\begin{minipage}[t]{8ex}{\color{red}\bf
\begin{verbatim}
(%i46) 
\end{verbatim}}
\end{minipage}
\begin{minipage}[t]{\textwidth}{\color{blue}
\begin{verbatim}
at(diff(f(f(x)),x),x=x[0]);
\end{verbatim}}
\end{minipage}
\definecolor{labelcolor}{RGB}{100,0,0}
\begin{math}\displaystyle
\parbox{8ex}{\color{labelcolor}(\%o46)}
f'\left( {x}_{0}\right) \,f'\left( {x}_{1}\right) 
\end{math}\\

\noindent It is then easy to prove (by induction) that
\[
\frac{\mathrm{d}f_N}{\mathrm{d}x} (x_0)=\prod^{N-1}_{i=0}f'(x_i),
\]
so
\begin{equation}\label{lyapunov}
\lambda (x_0)=\lim_{N\to\infty}\frac{1}{N}\log \left| \prod^{N-1}_{i=0}f'(x_i) \right|
			 =\lim_{N\to\infty}\frac{1}{N} \sum^{N-1}_{i=0} \log \left| f'(x_i) \right| .
\end{equation}

Using equation \eqref{lyapunov} we can compute the Lyapunov exponents numerically, approximating the limit
by a finite sum for $N$ big enough. As an example, consider the logistic map with parameter $r=3$. 
In (\%i8) we defined the function $F(r,x)$, and now we define its derivative:

\noindent
\begin{minipage}[t]{8ex}{\color{red}\bf
\begin{verbatim}
(%i48) 
\end{verbatim}}
\end{minipage}
\begin{minipage}[t]{\textwidth}{\color{blue}
\begin{verbatim}
define(dF(r,x),diff(F(r,x),x))$
\end{verbatim}}
\end{minipage}\\

\noindent Then we set the parameter value:

\noindent
\begin{minipage}[t]{8ex}{\color{red}\bf
\begin{verbatim}
(%i49) 
\end{verbatim}}
\end{minipage}
\begin{minipage}[t]{\textwidth}{\color{blue}
\begin{verbatim}
r:3$
\end{verbatim}}
\end{minipage}\\

\noindent Here we set the number of iterations:

\noindent
\begin{minipage}[t]{8ex}{\color{red}\bf
\begin{verbatim}
(%i50) 
\end{verbatim}}
\end{minipage}
\begin{minipage}[t]{\textwidth}{\color{blue}
\begin{verbatim}
maxiter:50000$
\end{verbatim}}
\end{minipage}\\

\noindent and the initial condition\footnote{We denote it by $x_1$ instead of $x_0$ because we are
going to make a list  with the iterations, with the initial condition as the first element and, for 
Maxima, lists are enumerated starting with the index $1$.}:

\noindent
\begin{minipage}[t]{8ex}{\color{red}\bf
\begin{verbatim}
(%i51) 
\end{verbatim}}
\end{minipage}
\begin{minipage}[t]{\textwidth}{\color{blue}
\begin{verbatim}
x[1]:random(1.0);
\end{verbatim}}
\end{minipage}
\definecolor{labelcolor}{RGB}{100,0,0}
\begin{math}\displaystyle
\parbox{8ex}{\color{labelcolor}(\%o51) }
{.}4823905248516196
\end{math}\\

\noindent Now, we construct the orbit from the initial condition:

\noindent
\begin{minipage}[t]{8ex}{\color{red}\bf
\begin{verbatim}
(%i52) 
\end{verbatim}}
\end{minipage}
\begin{minipage}[t]{\textwidth}{\color{blue}
\begin{verbatim}
for j:2 thru maxiter do (x[j]:F(r,x[j-1]))$
\end{verbatim}}
\end{minipage}

\noindent Finally, we estimate the corresponding Lyapunov exponent:

\noindent
\begin{minipage}[t]{8ex}{\color{red}\bf
\begin{verbatim}
(%i53) 
\end{verbatim}}
\end{minipage}
\begin{minipage}[t]{\textwidth}{\color{blue}
\begin{verbatim}
L:0$
\end{verbatim}}
\end{minipage}

\noindent
\begin{minipage}[t]{8ex}{\color{red}\bf
\begin{verbatim}
(%i54) 
\end{verbatim}}
\end{minipage}
\begin{minipage}[t]{\textwidth}{\color{blue}
\begin{verbatim}
for j:1 thru maxiter do L:L+log(abs(dF(r,x[j])))$
\end{verbatim}}
\end{minipage}

\noindent
\begin{minipage}[t]{8ex}{\color{red}\bf
\begin{verbatim}
(%i55) 
\end{verbatim}}
\end{minipage}
\begin{minipage}[t]{\textwidth}{\color{blue}
\begin{verbatim}
'%lambda=L/maxiter;
\end{verbatim}}
\end{minipage}
\definecolor{labelcolor}{RGB}{100,0,0}
\begin{math}\displaystyle
\parbox{8ex}{\color{labelcolor}(\%o55) }
\lambda=-3{.}145501884323275\,{10}^{-4}
\end{math}\\

\noindent For this case, $r=3$, we get $\lambda \simeq 0$. The reader can experimentally check that for
the logistic map $\lambda (x_0)=\lambda$, that is, the Lyapunov exponent does not depend on the initial
condition $x_0 \in ]0,1[$, it only depends on the parameter $r$. For values of $r$ lesser than
(approximately) $3{.}569945$, we have that $\lambda \leq 0$, but if $r> 3{.}569945$, the behavior of $\lambda$
as function of $r$ becomes quite complicated.

\section{The Duffing oscillator}
We have only considered systems described by first-order equations. However, most of the physical systems of interest
are described by second order equations. In this section we present and example of these, the Duffing oscillator.

Let us consider a general Newtonian system, described by a second-order differential equation:
\[
                            \ddot{x}(t)=F(x(t),\dot{x}(t))
\]
If $F=F(x)=-\mathrm{d}V/\mathrm{d}x$, we say that the system is conservative and the function $V=V(x)$
is called the \emph{potential}. For example, the system
\[
 \ddot{x}(t)=-\frac{1}{4}x^3(t)+x(t)
\]

\noindent is conservative and its potential is given by

\noindent
\begin{minipage}[t]{8ex}{\color{red}\bf
\begin{verbatim}
(%i56) 
\end{verbatim}}
\end{minipage}
\begin{minipage}[t]{\textwidth}{\color{blue}
\begin{verbatim}
'V(x)=integrate(x^3/4-x,x);
\end{verbatim}}
\end{minipage}
\definecolor{labelcolor}{RGB}{100,0,0}
\begin{math}\displaystyle
\parbox{8ex}{\color{labelcolor}(\%o56) }
\mathrm{V}\left( x\right) =\frac{{x}^{4}}{16}-\frac{{x}^{2}}{2}
\end{math}\\

The graph of this function is:

\noindent
\begin{minipage}[t]{8ex}{\color{red}\bf
\begin{verbatim}
(%i57) 
\end{verbatim}}
\end{minipage}
\begin{minipage}[t]{\textwidth}{\color{blue}
\begin{verbatim}
wxplot2d(integrate(x^3/4-x,x),[x,-3,3],[ylabel,"V(x)"]);
\end{verbatim}}
\end{minipage}
\begin{math}\displaystyle
\parbox{8ex}{\color{labelcolor}(\%t57) }
\includegraphics[width=8cm]{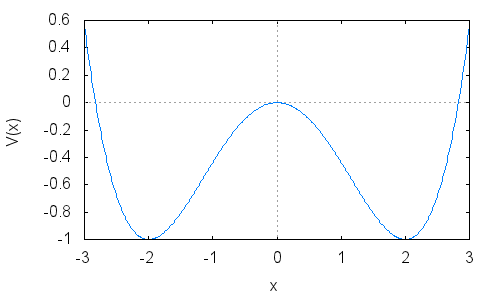}
\parbox{8ex}{\color{labelcolor}(\%o57) }
\end{math}\\

We observe an unstable equilibrium at $x=0$ and two stable equilibria at
$x=-2$ and $x=2$. Now let us modify the system by including a velocity-dependent \emph{damping term}:
\[
 \ddot{x}(t)=-\frac{1}{4}x^ 3(t)+x(t)-\frac{1}{10}\dot{x}(t)
\]

\noindent The system is no longer conservative, since the damping term dissipates energy
in the form of heat. As a result, the oscillations made by the system decrease in
amplitude until there is no motion. We illustrate this by computing the trajectory
with initial conditions $x(0)=3$, $\dot{x}(0)=10$:

\noindent
\begin{minipage}[t]{8ex}{\color{red}\bf
\begin{verbatim}
(%i58) 
\end{verbatim}}
\end{minipage}
\begin{minipage}[t]{\textwidth}{\color{blue}
\begin{verbatim}
duff:[v,-v/10+x-x^3/4]$
\end{verbatim}}
\end{minipage}

\noindent
\begin{minipage}[t]{8ex}{\color{red}\bf
\begin{verbatim}
(%i59) 
\end{verbatim}}
\end{minipage}
\begin{minipage}[t]{\textwidth}{\color{blue}
\begin{verbatim}
icduff:[3,10]$
\end{verbatim}}
\end{minipage}

\noindent
\begin{minipage}[t]{8ex}{\color{red}\bf
\begin{verbatim}
(%i60) 
\end{verbatim}}
\end{minipage}
\begin{minipage}[t]{\textwidth}{\color{blue}
\begin{verbatim}
solduff:rk(duff,[x,v],icduff,[t,0,100,0.1])$
\end{verbatim}}
\end{minipage}

\noindent
\begin{minipage}[t]{8ex}{\color{red}\bf
\begin{verbatim}
(%i61) 
\end{verbatim}}
\end{minipage}
\begin{minipage}[t]{\textwidth}{\color{blue}
\begin{verbatim}
curveduff:map(lambda([x],rest(x,-1)),solduff)$
\end{verbatim}}
\end{minipage}

\noindent
\begin{minipage}[t]{8ex}{\color{red}\bf
\begin{verbatim}
(%i62) 
\end{verbatim}}
\end{minipage}
\begin{minipage}[t]{\textwidth}{\color{blue}
\begin{verbatim}
pointsduff:map(lambda([x],rest(x)),solduff)$
\end{verbatim}}
\end{minipage}

\noindent
\begin{minipage}[t]{8ex}{\color{red}\bf
\begin{verbatim}
(%i63) 
\end{verbatim}}
\end{minipage}
\begin{minipage}[t]{\textwidth}{\color{blue}
\begin{verbatim}
wxdraw2d(point_type=none,points_joined=true,
xlabel="t",ylabel="x(t)",points(curveduff));
\end{verbatim}}
\end{minipage}
\begin{math}\displaystyle
\parbox{8ex}{\color{labelcolor}(\%t63) }
\includegraphics[width=8cm]{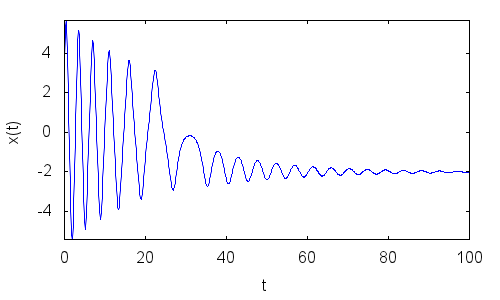}
\parbox{8ex}{\color{labelcolor}(\%o63) }
\end{math}\\

We can also represent the dynamics of the system in the plane
$(x,v=\mathrm{d}x/\mathrm{d}t)$ (\emph{the phase plane}):

\noindent
\begin{minipage}[t]{8ex}{\color{red}\bf
\begin{verbatim}
(%i64) 
\end{verbatim}}
\end{minipage}
\begin{minipage}[t]{\textwidth}{\color{blue}
\begin{verbatim}
wxdraw2d(point_type=none,points_joined=true,
xlabel="x(t)",ylabel="v(t)",points(pointsduff));
\end{verbatim}}
\end{minipage}
\begin{math}\displaystyle
\parbox{8ex}{\color{labelcolor}(\%t64) }
\includegraphics[width=8cm]{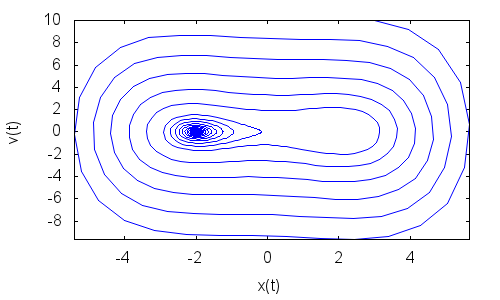}
\parbox{8ex}{\color{labelcolor}(\%o64) }
\end{math}\\

\noindent Note that this result coincides with (\%o63), where it is clear
that the system evolves to the stable equilibrium $x=-2$, oscillating around this value
with decreasing amplitude until it stops.

The \emph{Duffing oscillator} is obtained including in the previous system and
oscillating force (the \emph{forcing term}) $\sin(\omega t)$:
\[
 \ddot{x}(t)=-\frac{1}{4}x^3 (t)+x(t)-\frac{1}{10}\dot{x}(t)+\sin (\omega t)
\]

\noindent The effect of the forcing term is to restore the energy dissipated by
the damping term. If the frequency of the forcing term is the appropriated one, the system
will oscillate in a stable way, but if there is no synchronization with respect to the
non-forced system, a chaotic behavior will appear. For example, for $\omega =1$
starting from the unstable equilibrium ($x=0,v=0$), after
a short transient the system will evolve to a stable oscillating regime (a \emph{limit cycle}):

\noindent
\begin{minipage}[t]{8ex}{\color{red}\bf
\begin{verbatim}
(%i65) 
\end{verbatim}}
\end{minipage}
\begin{minipage}[t]{\textwidth}{\color{blue}
\begin{verbatim}
duffing:[v,-v/10+x-x^3/4+sin(t)]$
\end{verbatim}}
\end{minipage}

\noindent
\begin{minipage}[t]{8ex}{\color{red}\bf
\begin{verbatim}
(%i66) 
\end{verbatim}}
\end{minipage}
\begin{minipage}[t]{\textwidth}{\color{blue}
\begin{verbatim}
icduffing:[0,0]$
\end{verbatim}}
\end{minipage}

\noindent
\begin{minipage}[t]{8ex}{\color{red}\bf
\begin{verbatim}
(%i67) 
\end{verbatim}}
\end{minipage}
\begin{minipage}[t]{\textwidth}{\color{blue}
\begin{verbatim}
sduffing:rk(duffing,[x,v],icduffing,[t,0,100,0.1])$
\end{verbatim}}
\end{minipage}

\noindent
\begin{minipage}[t]{8ex}{\color{red}\bf
\begin{verbatim}
(%i68) 
\end{verbatim}}
\end{minipage}
\begin{minipage}[t]{\textwidth}{\color{blue}
\begin{verbatim}
cduffing:map(lambda([x],rest(x,-1)),sduffing)$
\end{verbatim}}
\end{minipage}

\noindent
\begin{minipage}[t]{8ex}{\color{red}\bf
\begin{verbatim}
(%i69) 
\end{verbatim}}
\end{minipage}
\begin{minipage}[t]{\textwidth}{\color{blue}
\begin{verbatim}
pduffing:map(lambda([x],rest(x)),sduffing)$
\end{verbatim}}
\end{minipage}

\noindent
\begin{minipage}[t]{8ex}{\color{red}\bf
\begin{verbatim}
(%i70) 
\end{verbatim}}
\end{minipage}
\begin{minipage}[t]{\textwidth}{\color{blue}
\begin{verbatim}
wxdraw2d(point_type=none,points_joined=true,color=magenta,
xlabel="t",ylabel="x(t)",points(cduffing));
\end{verbatim}}
\end{minipage}
\begin{math}\displaystyle
\parbox{8ex}{\color{labelcolor}(\%t70) }
\includegraphics[width=8cm]{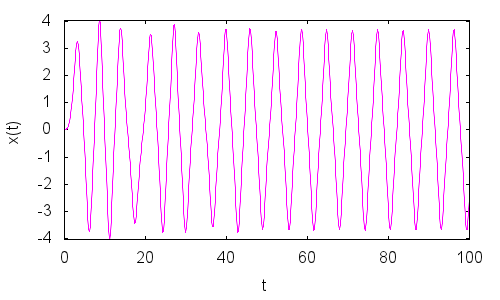}
\parbox{8ex}{\color{labelcolor}(\%o70) }
\end{math}\\

The corresponding phase plane diagram shows that the system tends to a
limit cycle, oscillating around the two stable equilibria, instead of
approaching just one of them, as happens for the non-forced case:

\noindent
\begin{minipage}[t]{8ex}{\color{red}\bf
\begin{verbatim}
(%i71) 
\end{verbatim}}
\end{minipage}
\begin{minipage}[t]{\textwidth}{\color{blue}
\begin{verbatim}
wxdraw2d(point_type=none,points_joined=true,color=magenta,
xlabel="x(t)",ylabel="v(t)",points(pduffing));
\end{verbatim}}
\end{minipage}
\begin{math}\displaystyle
\parbox{8ex}{\color{labelcolor}(\%t71) }
\includegraphics[width=8.25cm]{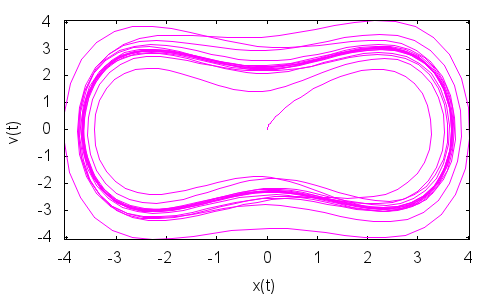}
\parbox{8ex}{\color{labelcolor}(\%o71) }
\end{math}\\

The behaviors considered do not exhaust the possibilities for the dynamics of the Duffing oscillator.
In the next section we will see how to analyze the chaotic case.

\section{Poincar\'e sections}
In general, for a second order differential equation of the form
\[
                    \ddot{x}(t)=F(t,x,\dot{x})
\]
or, equivalently (introducing the variables $u(t)=x(t)$, $v(t)=\dot{x}(t)$) for the first order system
\[
\begin{cases}
\displaystyle \frac{\mathrm{d}u}{\mathrm{d}t}=v \\
 \\
\displaystyle \frac{\mathrm{d}v}{\mathrm{d}t}=F(t,u,v)
\end{cases},
\]
the trajectories in the phase plane can intersect themselves and the resulting diagram becomes quite complicated.
Let us note that this is not a contradiction with the theorem on the uniqueness of solutions, because here we have
a non autonomous differential equation and the phase plane is obtained projecting from the phase space
$(t,x,v)\mapsto (x,v)$. The theorem applies, when it is the case, in the variables  $(t,x,v)$.
For example, for the Duffing oscillator with forcing term  $2{.}5\sin (2t)$, we get the following phase diagram:

\noindent
\begin{minipage}[t]{8ex}{\color{red}\bf
\begin{verbatim}
(%i72) 
\end{verbatim}}
\end{minipage}
\begin{minipage}[t]{\textwidth}{\color{blue}
\begin{verbatim}
duffing1:[v,-v/10+x-x^3/4+2.5*sin(2*t)]$
\end{verbatim}}
\end{minipage}

\noindent
\begin{minipage}[t]{8ex}{\color{red}\bf
\begin{verbatim}
(%i73) 
\end{verbatim}}
\end{minipage}
\begin{minipage}[t]{\textwidth}{\color{blue}
\begin{verbatim}
icduffing1:[0,0]$
\end{verbatim}}
\end{minipage}

\noindent
\begin{minipage}[t]{8ex}{\color{red}\bf
\begin{verbatim}
(%i74) 
\end{verbatim}}
\end{minipage}
\begin{minipage}[t]{\textwidth}{\color{blue}
\begin{verbatim}
sduffing1:rk(duffing1,[x,v],icduffing1,[t,0,100,0.1])$
\end{verbatim}}
\end{minipage}

\noindent
\begin{minipage}[t]{8ex}{\color{red}\bf
\begin{verbatim}
(%i75) 
\end{verbatim}}
\end{minipage}
\begin{minipage}[t]{\textwidth}{\color{blue}
\begin{verbatim}
pduffing1:map(lambda([x],rest(x)),sduffing1)$
\end{verbatim}}
\end{minipage}

\noindent
\begin{minipage}[t]{8ex}{\color{red}\bf
\begin{verbatim}
(%i76) 
\end{verbatim}}
\end{minipage}
\begin{minipage}[t]{\textwidth}{\color{blue}
\begin{verbatim}
wxdraw2d(point_type=none,points_joined=true,color=coral,
xlabel="x(t)",ylabel="v(t)",points(pduffing1));
\end{verbatim}}
\end{minipage}
\begin{math}\displaystyle
\parbox{8ex}{\color{labelcolor}(\%t76) }
\includegraphics[width=7.75cm]{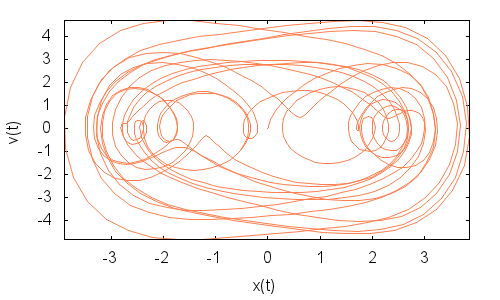}
\parbox{8ex}{\color{labelcolor}(\%o76) }
\end{math}\\

Poincar\'e introduced a very useful technique to study the dynamics in these situations. It
consists in reckoning a trajectory in the phase space $(x,v,t)$ by using
hyperplanes equally spaced in time as ``check points'', and then projecting the resulting
points on the plane $(x,v)$:

\begin{figure}[h]
\label{figura1}
  \begin{center}
    \includegraphics[scale=0.35]{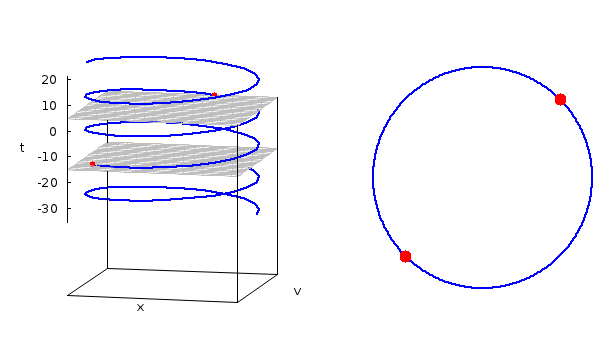}
  \end{center}
  \caption{Poincar\'e sections}
\end{figure}

Of course, if the time interval between these hyperplanes (called the \emph{Poincar\'e stroboscopic sections})
is $T$ and the system is periodic with period $T$, we will see just a point on the phase plane.
In the figure we see a quasi--periodic system: on the phase plane the trajectories are
circumferences, but several points can appear in the sections projection, showing that the system
oscillates between several equilibria with a frequency different from the time interval between
the Poincar\'e sections.

When the dynamics is chaotic, the diagram obtained by projecting the Poincar\'e sections can be
really complicated, usually displaying fractal structure.\\
Let us compute the Poincar\'e sections for the case of the Duffing oscillator. First, we consider the case
of a period $T=\frac{2\pi}{\omega}$ corresponding to the frequency $\omega =1$ (that is, the
forcing term is $\sin (t)$). We register the points in the orbit at time intervals of length
 $T=\frac{2\pi}{\omega}=2\pi$:

\noindent
\begin{minipage}[t]{8ex}{\color{red}\bf
\begin{verbatim}
(%i77) 
\end{verbatim}}
\end{minipage}
\begin{minipage}[t]{\textwidth}{\color{blue}
\begin{verbatim}
miter:25$
\end{verbatim}}
\end{minipage}

\noindent
\begin{minipage}[t]{8ex}{\color{red}\bf
\begin{verbatim}
(%i78) 
\end{verbatim}}
\end{minipage}
\begin{minipage}[t]{\textwidth}{\color{blue}
\begin{verbatim}
%tau:bfloat(%pi)$
\end{verbatim}}
\end{minipage}

\noindent
\begin{minipage}[t]{8ex}{\color{red}\bf
\begin{verbatim}
(%i79) 
\end{verbatim}}
\end{minipage}
\begin{minipage}[t]{\textwidth}{\color{blue}
\begin{verbatim}
sduffing2:rk(duffing,[x,v],icduffing,
[t,0,miter*2*%tau,%tau/30])$
\end{verbatim}}
\end{minipage}

\noindent
\begin{minipage}[t]{8ex}{\color{red}\bf
\begin{verbatim}
(%i80) 
\end{verbatim}}
\end{minipage}
\begin{minipage}[t]{\textwidth}{\color{blue}
\begin{verbatim}
pduffing2:create_list(sduffing2[i],i,makelist(i*60,i,1,miter))$
\end{verbatim}}
\end{minipage}

\noindent
\begin{minipage}[t]{8ex}{\color{red}\bf
\begin{verbatim}
(%i81) 
\end{verbatim}}
\end{minipage}
\begin{minipage}[t]{\textwidth}{\color{blue}
\begin{verbatim}
poinduffing2:map(lambda([x],rest(x)),
makelist(pduffing2[i],i,1,miter))$
\end{verbatim}}
\end{minipage}

\noindent
\begin{minipage}[t]{8ex}{\color{red}\bf
\begin{verbatim}
(%i82) 
\end{verbatim}}
\end{minipage}
\begin{minipage}[t]{\textwidth}{\color{blue}
\begin{verbatim}
wxdraw2d(point_type=filled_circle,color=magenta,xtics=1,
ytics=1,xrange=[-4,1.5],yrange=[0,4],points(poinduffing2));
\end{verbatim}}
\end{minipage}
\begin{math}\displaystyle
\parbox{8ex}{\color{labelcolor}(\%t82) }
\includegraphics[width=8cm]{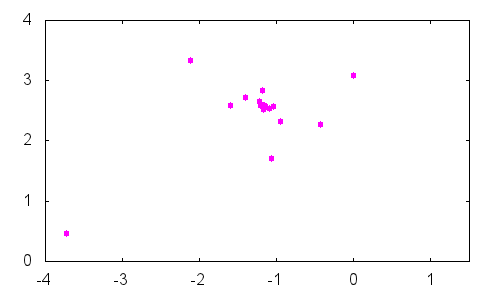}
\displaystyle
\parbox{8ex}{\color{labelcolor}(\%o82) }
\end{math}\\

With the exception of some sparse points representing the transient, we see that
all the points tend to a cluster (oscillations occur close to a limit cycle). To
show that the most distant points corresponds to the transient, we eliminate, for example,
the first four samples (corresponding to the sections at $t=0,2\pi,4\pi$ and $6\pi$):

\noindent
\begin{minipage}[t]{8ex}{\color{red}\bf
\begin{verbatim}
(%i83) 
\end{verbatim}}
\end{minipage}
\begin{minipage}[t]{\textwidth}{\color{blue}
\begin{verbatim}
transient:map(lambda([x],rest(x)),
makelist(pduffing2[i],i,5,miter))$
\end{verbatim}}
\end{minipage}

\noindent
\begin{minipage}[t]{8ex}{\color{red}\bf
\begin{verbatim}
(%i84) 
\end{verbatim}}
\end{minipage}
\begin{minipage}[t]{\textwidth}{\color{blue}
\begin{verbatim}
wxdraw2d(point_type=filled_circle,color=magenta,xtics=1,
ytics=1,xrange=[-4,1.5],yrange=[0,4],points(transient));
\end{verbatim}}
\end{minipage}
\begin{math}\displaystyle
\parbox{8ex}{\color{labelcolor}(\%t84) }
\includegraphics[width=8cm]{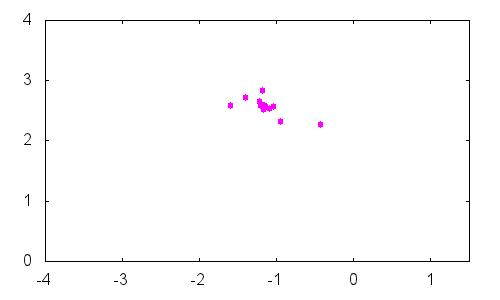}
\displaystyle
\parbox{8ex}{\color{labelcolor}(\%o84) }
\end{math}\\

Of course, increasing the number of samples we get a more illustrative picture. But the computational effort
is not worth in this case. There are more interesting situations, as it is the case of the Duffing oscillator
with forcing term $2{.}5\sin (2t)$. Now the period corresponds to the frequency $\omega =2$, that is,
$T=\frac{2\pi}{2}=\pi$. We let the system evolve for a long time in order to get a large
sample via the Poincar\'e sections:

\noindent
\begin{minipage}[t]{8ex}{\color{red}\bf
\begin{verbatim}
(%i85) 
\end{verbatim}}
\end{minipage}
\begin{minipage}[t]{\textwidth}{\color{blue}
\begin{verbatim}
duffing3:[v,-v/10+x-x^3/4+2.5*sin(2*t)]$
\end{verbatim}}
\end{minipage}

\noindent
\begin{minipage}[t]{8ex}{\color{red}\bf
\begin{verbatim}
(%i86) 
\end{verbatim}}
\end{minipage}
\begin{minipage}[t]{\textwidth}{\color{blue}
\begin{verbatim}
icduffing3:[0,0]$
\end{verbatim}}
\end{minipage}

\noindent
\begin{minipage}[t]{8ex}{\color{red}\bf
\begin{verbatim}
(%i87) 
\end{verbatim}}
\end{minipage}
\begin{minipage}[t]{\textwidth}{\color{blue}
\begin{verbatim}
maxiter:1000$
\end{verbatim}}
\end{minipage}

\noindent
\begin{minipage}[t]{8ex}{\color{red}\bf
\begin{verbatim}
(%i88) 
\end{verbatim}}
\end{minipage}
\begin{minipage}[t]{\textwidth}{\color{blue}
\begin{verbatim}
sduffing3:rk(duffing3,[x,v],icduffing3,
[t,0,maxiter*%tau,%tau/30])$
\end{verbatim}}
\end{minipage}

\noindent
\begin{minipage}[t]{8ex}{\color{red}\bf
\begin{verbatim}
(%i89) 
\end{verbatim}}
\end{minipage}
\begin{minipage}[t]{\textwidth}{\color{blue}
\begin{verbatim}
pduffing3:
create_list(sduffing3[i],i,makelist(i*30,i,1,maxiter))$
\end{verbatim}}
\end{minipage}

\noindent
\begin{minipage}[t]{8ex}{\color{red}\bf
\begin{verbatim}
(%i90) 
\end{verbatim}}
\end{minipage}
\begin{minipage}[t]{\textwidth}{\color{blue}
\begin{verbatim}
poinduffing3:map(lambda([x],rest(x)),
makelist(pduffing3[i],i,1,maxiter))$
\end{verbatim}}
\end{minipage}

\noindent
\begin{minipage}[t]{8ex}{\color{red}\bf
\begin{verbatim}
(%i91) 
\end{verbatim}}
\end{minipage}
\begin{minipage}[t]{\textwidth}{\color{blue}
\begin{verbatim}
wxdraw2d(point_size=0.3,point_type=circle,
xrange=[-5,5],yrange=[-7,3],xtics=1,ytics=1,
color=coral,points(poinduffing3),grid=true);
\end{verbatim}}
\end{minipage}
\begin{math}\displaystyle
\parbox{8ex}{\color{labelcolor}(\%t91) }
\includegraphics[width=9cm]{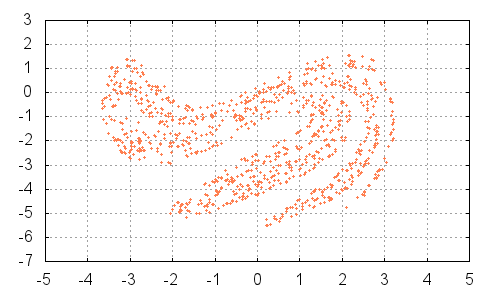}
\displaystyle
\parbox{8ex}{\color{labelcolor}(\%o91) }
\end{math}\\

\section{Fractal dimension}\label{fractals}
As we have pointed out in previous sections, the appearance of \emph{strange attractors}
is a usual feature of chaotic systems. These are sets around which the orbits move asymptotically
(thus justifying the name of attractor) and frequently have fractal dimension. We will introduce a
particular notion of fractal dimension (there are other possibilities) in the case of a
two dimensional dynamical system, so the attracting set $A$ will be a subset of
the plane. An example of this situation is given by the Duffing oscillator. The \emph{box-counting dimension}
of $A$ is defined as follows: we start with a square of side length $l$ enclosing $A$. In the next step, we divide
the square in $r$ sub--squares, each having side length $l/r$, and so on. In the $k$--step we have a grid with
$r^{k-1}$ squares, each with side length $\delta_k =\frac{l}{r^{k-1}}$, that cover the set $A$ as in (\%o91).
For each step $k$, let $N(k)$ be the number of squares containing at least \emph{some} point of $A$.\\
Let us observe that if $A$ were a smooth curve (and, therefore, homeomorphic to a segment of $\mathbb{R}$, to which we
would assign a topological dimension $1$), for a small enough value of the side of the squares each of them would contain
a piece of the curve with length as close as we wish to the side length of the square. So, in the case of a smooth
curve, we would have
\[
\lim_{k\to\infty} N(k)\cdot \delta_k = L,
\]
with $L$ the length of the curve. In this case we say that the scale of $N(k)$ goes as $\delta^{-1}_k$,
that is, $N(k)\simeq \delta^{-1}_k$. Analogously, if $A$ were a measurable set on the plane
(a set having an area and, thus, of dimension $2$), we would have
$N(k)\simeq \delta^{-2}_k$. Let us note that in these cases, that can be called ``regular'' ones,
the exponent $D$ in the relation $N(k)\simeq \delta^{-D}_k$ equals the topological dimension of the set.
Then, a \emph{fractal set} can be defined as a set for which a relation of the kind $N(k)\simeq \delta^{-D}_k$ holds,
with $D$ a positive real number (not necessarily an integer). In general, we call this number $D$ the \emph{box counting dimension} of the set, and remar that it satisfies
\[
D=\lim_{k\to\infty}\frac{\log N(k)}{-\log \delta_k}.
\]
This is a constructive definition that suggest how to perform the calculation of the box-counting dimension for a given set. Explicitly, we can plot $\log N(k)$ versus $-\log \delta_k$, then we can fit the data and finally
estimate $D$ as the slope of the regression line. As an example, let us calculate the dimension of the
Duffing attractor (\%i85) with this procedure.
We note that the Duffing attractor is contained in the box $[-5,5]\times [-7,3]$. We divide this box in
$10\times 10$, $20\times 20$,..., $200\times 200$ boxes, counting how many points are inside them.
As a detail, we first ``normalize'' the coordinates in such a way that they lie in the box $[0,10]\times [0,10]$.
The algorithm starts with a matrix filled with zeros and changes the scale at successive steps multiplying these lengths by $2,3,...,20$:

\noindent
\begin{minipage}[t]{8ex}{\color{red}\bf
\begin{verbatim}
(%i92) 
\end{verbatim}}
\end{minipage}
\begin{minipage}[t]{\textwidth}{\color{blue}
\begin{verbatim}
resolution:20$
\end{verbatim}}
\end{minipage}

\noindent
\begin{minipage}[t]{8ex}{\color{red}\bf
\begin{verbatim}
(%i93) 
\end{verbatim}}
\end{minipage}
\begin{minipage}[t]{\textwidth}{\color{blue}
\begin{verbatim}
for n:1 thru resolution do
 (Z[n]:substpart("[",zeromatrix(10*n,10*n),0),
  boxcount[n]:0,
  for k:1 thru maxiter do (
   ix:floor(n*(poinduffing3[k][1]+5)),
   iy:floor(n*(poinduffing3[k][2]+7)),
   if is(Z[n][ix][iy]=0) then 
    (Z[n][ix][iy]:1,boxcount[n]:boxcount[n]+1)))$
\end{verbatim}}
\end{minipage}\\

\noindent Here is the plot of the pairs of numbers obtained.

\noindent
\begin{minipage}[t]{8ex}{\color{red}\bf
\begin{verbatim}
(%i94) 
\end{verbatim}}
\end{minipage}
\begin{minipage}[t]{\textwidth}{\color{blue}
\begin{verbatim}
fitdim:makelist(
[log(n)/log(10),log(boxcount[n])],n,1,resolution),numer$
\end{verbatim}}
\end{minipage}

\noindent
\begin{minipage}[t]{9ex}{\color{red}\bf
\begin{verbatim}
(%i95) 
\end{verbatim}}
\end{minipage}
\begin{minipage}[t]{\textwidth}{\color{blue}
\begin{verbatim}
wxdraw2d(point_size=1,point_type=filled_circle,
color=dark_violet,points(fitdim));
\end{verbatim}}
\end{minipage}
\begin{math}\displaystyle
\parbox{9ex}{\color{labelcolor}(\%t95) }
\includegraphics[width=8.25cm]{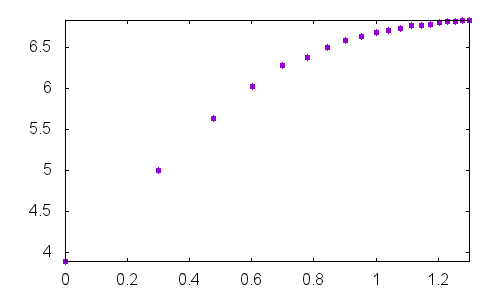}
\displaystyle
\parbox{9ex}{\color{labelcolor}(\%o95) }
\end{math}\\

\noindent The first values are ``outliers'' (since the boxes are in fact too big), so we can safely discard them.
In the same way, for the last values the boxes are too small, counting nearly one point per box, so again we can
discard them. This situation, of data that gives a wrong contribution to the estimation of the box-counting dimension,
is a well known problem discussed in the literature (see \cite{Kli 94}); in fact, there are several algorithms
aimed to automatically selecting the most representative data, but for our (limited) purposes, we just use the
visual information in the graph (\%o95). According with this, it seems a good choice to eliminate the first 7 
values and the two last ones to make the fitting. The result is:

\noindent
\begin{minipage}[t]{9ex}{\color{red}\bf
\begin{verbatim}
(%i96) 
\end{verbatim}}
\end{minipage}
\begin{minipage}[t]{\textwidth}{\color{blue}
\begin{verbatim}
fitdimension:rest(rest(fitdim,7),-2)$
\end{verbatim}}
\end{minipage}

\noindent
\begin{minipage}[t]{9ex}{\color{red}\bf
\begin{verbatim}
(%i97) 
\end{verbatim}}
\end{minipage}
\begin{minipage}[t]{\textwidth}{\color{blue}
\begin{verbatim}
load(stats)$
\end{verbatim}}
\end{minipage}

\noindent
\begin{minipage}[t]{9ex}{\color{red}\bf
\begin{verbatim}
(%i98) 
\end{verbatim}}
\end{minipage}
\begin{minipage}[t]{\textwidth}{\color{blue}
\begin{verbatim}
simple_linear_regression(fitdimension)$
\end{verbatim}}
\end{minipage}

\noindent
\begin{minipage}[t]{9ex}{\color{red}\bf
\begin{verbatim}
(%i99) 
\end{verbatim}}
\end{minipage}
\begin{minipage}[t]{\textwidth}{\color{blue}
\begin{verbatim}
fiteqn:take_inference(
model,simple_linear_regression(fitdimension));
\end{verbatim}}
\end{minipage}
\definecolor{labelcolor}{RGB}{100,0,0}
\begin{math}\displaystyle
\parbox{9ex}{\color{labelcolor}(\%o99)}
{.}6534032420835236\,x+6{.}021481801936183
\end{math}

\noindent
\begin{minipage}[t]{9ex}{\color{red}\bf
\begin{verbatim}
(%i100) 
\end{verbatim}}
\end{minipage}
\begin{minipage}[t]{\textwidth}{\color{blue}
\begin{verbatim}
fractaldimension:coeff(fiteqn,x);
\end{verbatim}}
\end{minipage}
\definecolor{labelcolor}{RGB}{100,0,0}
\begin{math}\displaystyle
\parbox{9ex}{\color{labelcolor}(\%o100)}
{.}6534032420835236
\end{math}\\

\noindent Since our approximation is very rough, we can round-off the value so obtained.
Therefore, we can say that the fractal dimension of the Duffing attractor is approximately $0{.}65$ (between $0$
and $1$, as it should be expected).

\end{document}